\documentclass[aps,prb,twocolumn,showpacs,nofootinbib]{revtex4-1}
\usepackage{amsfonts}
\usepackage{amscd}
\usepackage{amsmath}
\usepackage{amssymb}
\usepackage{txfonts}

\usepackage{graphicx}
\usepackage{dcolumn}
\usepackage{bm}
\usepackage{xcolor}
\usepackage[%
  colorlinks=true,
  urlcolor=blue,
  linkcolor=blue,
  citecolor=blue
]{hyperref}
\usepackage{tabularx}
\usepackage{etoolbox}
\newcommand{\nn}{\nonumber}
\newcommand{\beq}{\begin{eqnarray}}
\newcommand{\eeq}{\end{eqnarray}}

\def\so2{{\tt SO(2)}}

\def\d3d{{\tt D_{3d}}}

\def\c2v{{\tt C_{2v}}}
\makeatletter
\let\cat@comma@active\@empty
\makeatother
\begin{document}

\title{
Torsional chiral magnetic effect due to skyrmion textures in a Weyl superfluid $^3$He-A
}


\author{Yusuke Ishihara}
\author{Takeshi Mizushima}
\author{Atsushi Tsuruta}
\author{Satoshi Fujimoto}
\affiliation{Department of Materials Engineering Science, Osaka University, Toyonaka 560-8531, Japan }




\date{\today}

\begin{abstract}
We investigate torsional chiral magnetic effect (TCME) induced by skyrmion-vortex textures in the A phase of the superfluid $^3$He. In $^3$He-A, Bogoliubov quasiparticles around point nodes behave as Weyl fermions, and the nodal direction represented by the $\ell$-vector may form a spatially modulated texture. $\ell$-textures generate a chiral gauge field and a torsion field directly acting on the chirality of Weyl-Bogoliubov quasiparticles. It has been clarified by G. E. Volovik [Pi'sma Zh. Eksp. Teor. Fiz. {\bf 43}, 428 (1986)] that, if the $\ell$-vector is twisted as $\hat{\bm \ell}\!\cdot\!({\rm curl}\hat{\bm \ell})\!\neq\! 0$, the chiral gauge field is responsible for the chiral anomaly, leading to an anomalous current along ${\bm \ell}$. Here we show that, even if $\hat{\bm \ell}\!\cdot\!({\rm curl}\hat{\bm \ell})\!=\! 0$, a torsion arising from $\ell$-textures brings about contributions to the equilibrium currents of Weyl-Bogoliubov quasiparticles along ${\rm curl}{\bm \ell}$. This implies that while the anomalous current appears only for the twisted (Bloch-type) skyrmion of the $\ell$-vector, the extra mass current due to TCME always exists regardless of the skyrmion type. Solving the Bogoliubov-de Gennes equation, we demonstrate that both Bloch-type and N\'{e}el-type skyrmions induce chiral fermion states with spectral asymmetry, and possess spatially inhomogeneous structures of Weyl bands in the real coordinate space. Furthermore, we discuss the contributions of Weyl-Bogoliubov quasiparticles and continuum states to the mass current density in the vicinity of the topological phase transition. In the weak coupling limit, continuum states give rise to backflow to the mass current generated by Weyl-Bogoliubov quasiparticles, which makes a non-negligible contribution to the orbital angular momentum. As the topological transition is approached, the mass current density is governed by the contribution of continuum states.

%
%
\end{abstract}



\maketitle

\section{Introduction}
Weyl semimetals have been attracting much attention because of the realization of chiral anomaly in condensed matter systems, which is experimentally detectable
in various exotic transport phenomena such as the anomalous Hall effect, chiral magnetic effect, and negative magnetoresistance.\cite{NN,PhysRevB.83.205101,PhysRevLett.107.127205,PhysRevB.85.035103,1367-2630-9-9-356,zyuzinPRB12,goswamiPRB13,sonPRB13,liuPRB13,vazifehPRL13,hosur,vafek14,goswamiPRB15,yamamotoPRD15}
Chiral anomaly is the violation of conservation law of axial currents in the case with both electric and magnetic fields which are not orthogonal to each other.
Its origin is attributed to monopole charge carried by Weyl points in the momentum space, which generate the Berry curvature, and
are sources and drains of momentum generation.  
Recent experimental studies revealed the realization of chiral anomaly in Weyl semimetal materials via the observation of
negative magnetoresistance.\cite{MRreview,expMR1,expMR2} 

The notion of Weyl semimetals is naturally generalized to superconducting states.\cite{balents}
In superconductors with broken time-reversal symmetry such as chiral pairing states and non-unitary odd-parity pairing states, nodal excitations from point-nodes of
the superconducting gap behave as Weyl fermions accompanying the Berry curvature.
There are several candidate systems of Weyl superconductors and superfluids such as the A-phase of the superfluid $^3$He,\cite{volovik86,volovik03,volovik16,mizushimaJPSJ16} URu$_2$Si$_2$,\cite{goswami13} the $B$-phase of UPt$_3$,\cite{goswami15} UCoGe,\cite{Sato-Fujimoto} and the B-phase of U$_{1-x}$Th$_x$Be$_{13}$.~\cite{shimizuPRB17,mizushimaPRB18,machidaJPSJ18}
Since the Bogoliubov quasiparticles are the superposition of electrons and holes, the usual coupling with electromagnetic fields
does not directly lead to chiral anomaly.
However, it is still possible that in Weyl superconductors and Weyl superfluids, emergent electromagnetic fields arising from spatially inhomogeneous textures of the superconducting order parameter and its dynamics give rise to
chiral anomaly phenomena.
As a matter of fact, in 1997, more than ten years before the invention of the notion of Weyl semimetals,
Bevan {\it et al.} observed momentum generation due to chiral anomaly in $^3$He-A with skyrmion textures of the $\ell$-vector field,\cite{bevan} which was motivated by pioneering theoretical works of Volovik and his collaborators.\cite{volovik81,volovik84,volovik85,volovik86,volovik86-2,balatskii87,volovik95} 
In the experiment,\cite{bevan} the chiral anomaly was detected via the measurement of an extra force on skyrmion-vortices .

In this paper, we consider another chiral anomaly effect which is referred to as the torsional chiral magnetic effect (TCME).
The TCME was originally proposed for magnetic Weyl semimetals with lattice dislocations.\cite{SF}
Lattice dislocations give rise to torsion fields which cause emergent magnetic fields acting on Weyl fermions,
and result in equilibrium currents flowing along the dislocation lines. 
The current induced by the torsion field is given by,
\begin{eqnarray}
\bm{J}^{\rm TCME}=\frac{ev_F\Lambda}{4\pi^2}\sum_{a=x,y,z}\bm{T}^a(p_{L a}-p_{R a}), \label{eq:TCME}
\end{eqnarray}
where $v_F$ is the Fermi velocity, $\Lambda$ is the momentum cutoff, 
$p_{L(R) a}$ ($a=x,y,z$) is the position of the Weyl point with left(right)-handed chirality in momentum space, and
\begin{eqnarray}
(\bm{T}^a)^{\mu}=\frac{\epsilon^{\mu\nu\lambda}}{2}T_{\nu\lambda}^a,
\end{eqnarray}
 where $T_{\nu\lambda}^a$ is torsion which can be realized in condensed matter systems
by topological defects such as lattice dislocation and a skyrmion texture of magnetic order.
In the case of superconductors, torsional magnetic fields arise from vortex textures of the superconducting order parameter or lattice strain, and hence, the negative thermal magnetoresistivity, that is, the anomalous enhancement of longitudinal thermal conductivity along the torsional magnetic field.~\cite{kobayashi18}

\begin{figure}[t]
\includegraphics[width=85mm]{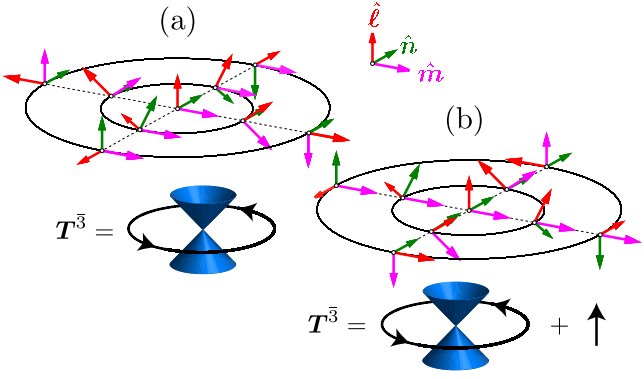}
\caption{Real-space skyrmion $\ell$-textures and ``torsional'' magnetic fields acting on Wely-Bogoliubov quasiparticles, ${\bm T}^{\bar{3}}={\rm curl}\hat{\bm \ell}$: (a) N\'{e}el-type skyrmion-vortex with $\alpha = 0$ and (b) Bloch-type skyrmion with $\alpha = \pi/2$ (see Eq.~\eqref{eq:ell}). {The former texture has ${\bm j}^{\rm TCME}\propto {\bm T}^{\bar{3}}$ and ${\bm j}^{\rm an}={\bm 0}$, while the latter has nonvanishing ${\bm j}^{\rm TCME}\propto {\bm T}^{\bar{3}}$ and ${\bm j}^{\rm an} \propto \hat{\bm \ell}$.}}
\label{fig:skyrmion}
\end{figure}

Here we focus on the A-phase of $^3$He having a skyrmion-vortex as a promising platform for the TCME. The order parameter tensor, $A_{\mu i}$, that transforms as a vector with respect to index $\mu=x,y,z$ ($i=x,y,z$) under spin (orbital) rotations, is given by the complex form~\cite{abm1,abm2}
\beq
A_{\mu i} = \Delta _{\rm A}(T) \hat{d}_{\mu} \left( \hat{\bm m}+ i\hat{\bm n}\right)_ie^{i\varphi},
\label{eq:abm}
\eeq
where $\hat{\bm d}$ and $(\hat{\bm m},\hat{\bm n})$ are unit vectors representing spin and orbital degrees of freedom in the superfluid vacuum, respectively. This is the Cooper-pair state with a definite orbital angular momentum represented by $\hat{\bm \ell}\equiv \hat{\bm m}\times \hat{\bm n}$, and the $\ell$-vector points to the nodal orientation at which Weyl-Bogoliubov quasiparticles reside. Owing to the spontaneously broken gauge-orbit symmetry, the rotation of the orbital part, $\hat{\bm m}+i\hat{\bm n}$, about $\hat{\bm \ell}$ is equivalent to the ${\rm U}(1)$ phase rotation $\varphi$. This implies that the superfluid current can be generated by the texture of the triad $(\hat{\bm m},\hat{\bm n},\hat{\bm \ell})$ without ${\rm U}(1)$ phase singularities. For rotating $^3$He-A, therefore, the $\ell$-vector field spontaneously forms a skyrmion-like texture as a ground state,\cite{salomaa,thuneberg,eltsov} which is known as the Anderson-Toulouse vortex and the Mermion-Ho vortex (see Fig.~\ref{fig:skyrmion}).\cite{VW,MH,AT,mizushimaJPSJ16} The $\ell$-texture fields also appear in $^3$He confined in a narrow cylinder.~\cite{wimanPRB15,wiman18} The relation between $\ell$-textures and the mass current density at zero temperatures was derived by Mermin and Muzikar~\cite{merminPRB80} as
\begin{align}
{\bm j}^{\rm MM}={\rho}{\bm v}_{{\rm s}}
+\frac{\hbar}{4M}{\rm curl}\rho\hat{\bm \ell}
-\frac{\hbar}{2M}C_0\hat{\bm \ell}(\hat{\bm \ell}\cdot{\rm curl}\hat{\bm \ell}),
\label{eq:currentMM}
\end{align}
where $\rho$ is the mass density of $^3$He atoms, $M$ is the mass, and $C_0\approx\rho$ in the weak coupling approximation. The first term in Eq.~\eqref{eq:currentMM} results from the superfluid velocity, and the second term arises from a variation of orbital angular momentum of the Cooper pair, $\hbar\hat{\bm \ell}$, which resembles the electric current induced by a variation of the magnetization density in materials. The third term is the anomalous current referred to as ${\bm j}^{\rm an}$, which brings about a longstanding issue involved with the McClure-Takagi paradox.~\cite{MT,ishikawaPTP80,merminPRB80,kitaJPSJ96,tsuruta} For twisted $\ell$ textures with $\hat{\bm \ell}\cdot {\rm curl}\hat{\bm \ell}\neq 0$, ${\bm j}^{\rm an}$ violates the McClure-Takagi relation that the ground state with an axially symmetric $\ell$-texture has the total angular momentum per particle $L_z/N=\hbar/2$. 

{Using the semiclassical theory for Weyl-Bogoliubov quasiparticles, we here consider the contribution of the TCME to equilibrium mass flow in $^3$He-A with skyrmion-vortices. The texture of the $\ell$ field brings both a chiral gauge field (${\bm A}^{\rm em}\propto \hat{\bm \ell}$) and torsion ($T^{a}_{\mu\nu}$) into Weyl-Bogoliubov quasiparticles. It has been discussed that, if $\hat{\bm \ell}$ is twisted as ${\bm \ell}\cdot({\rm curl}{\bm \ell})\neq 0$, the chiral gauge field directly acting on the chirality of Weyl fermions leads to an anomalous current, ${\bm j}^{\rm an}\propto\hat{\bm \ell}(\hat{\bm \ell}\cdot{\rm curl}\hat{\bm \ell})$,\cite{combescotPRB83,combescotPRB86,volovik81,volovik84,volovik85,volovik86,volovik86-2,balatskii87,volovik95} which is the chiral-anomaly due to the emergent gauge field ${\bm A}^{\rm em}$. Here we show that a ``torsion'' field, ${\bm T}^{\bar{3}}={\rm curl}\hat{\bm \ell}$, arising from a texture of the triad $(\hat{\bm m},\hat{\bm n},\hat{\bm \ell})$, gives rise to new torsional contributions as in Eq.~\eqref{eq:TCME}. This is the torsional-anomaly aspect of the second term of Eq.~\eqref{eq:currentMM}. While ${\bm j}^{\rm an}$ exists only for twisted (Bloch-type) skyrmions with $\hat{\bm \ell}\cdot{\rm curl}\hat{\bm \ell}\neq 0$ (Fig.~\ref{fig:skyrmion}(b)), ${\bm j}^{\rm TCME}\propto {\rm curl}\hat{\bm \ell}$ is always finite regardless of the skyrmion type (Fig.~\ref{fig:skyrmion}(a)).} 

{The Bogoliubov-de Gennes (BdG) equation enables a full quantum mechanical treatment of Bogoliubov quasiparticles and provides tractable and feasible approach to the vicinity of the topological phase transition. Solving the BdG equation, we demonstrate that N\'{e}el-type skyrmions induce chiral fermion states with spectral asymmetry, which are responsible for macroscopic mass flow along the azimuthal direction. The results are consistent with the semiclassical theory for Weyl-Bogoliubov quasiparticles with the torsional magnetic field, ${\bm T}^{\bar{3}}={\rm curl}\hat{\bm \ell}$. We also discuss the chiral fermion states and mass flow for both N\'{e}el-type and  Bloch-type skyrmions in the light of the torsional magnetic field and discrete symmetries prevented by the skyrmion-vortex. Furthermore, we show that skyrmion $\ell$-textures give rise to spatially inhomogeneous structures of the Weyl fermion band in addition to the torsional magnetic field\cite{takashima}; the position of Weyl points in the momentum space exhibits inhomogeneous textures in the real space. Lastly we discuss the contribution of the Weyl-Bogoliubov quasiparticles and continuum states to the mass current in the vicinity of the topological phase transition. In the weak coupling limit, continuum states give rise to backflow to the mass current generated by Weyl-Bogoliubov quasiparticles, which makes a non-negligible contribution to the orbital angular momentum. As the topological transition is approached, the mass current density is governed by the contribution of continuum states.}

The organization of this paper is as follows. In Sec.~\ref{sec:TCME}, we present semiclassical analysis for TCME in the case of Weyl superconductors/superfluids. In Sec.~\ref{sec:chiral}, we describe the numerical method for solving the Bogoliubov-de-Gennes (BdG) equation for our purpose, and show the intrinsic features of Weyl-Bogoliubov quasiparticles in the presence of the $\ell$-texture, such as spectral asymmetry and spatially inhomogeneous structures of the Weyl fermion band. In Sec.~\ref{sec:current}, we discuss the mass current density induced by the skyrmion texture from the viewpoint of the TCME.
The final section is devoted to conclusion and discussion.

\section{Torsional chiral magnetic effect and semiclassical analysis}
\label{sec:TCME}

\subsection{Semiclassical equation of motion for Weyl-Bogoliubov quasiparticles}

We consider the Bogoliubov-de Gennes (BdG) Hamiltonian for the Bogoliubov quasiparticles in the A-phase of the superfluid $^3$He, i.e., a three-dimensional (3D) chiral $p+ip$ superfluid, with spatially varying gap structures such as skyrmion-like $\ell$-vector textures of the Anderson-Toulouse vortex and the Mermin-Ho vortex.
We apply the path integral formulation in a curved space with nonzero torsion which is induced by a vortex structure.
In the Lagrangian in the Feynmann kernel, the spatially varying gap function is expressed as
$A_{\mu i}{p}_i/p_{\rm F}$ with the tensorial field $A_{\mu i}$ in Eq.~\eqref{eq:abm}. 
Here $\hat{\bm m}$ and $\hat{\bm n}$ are unit vectors for a local orthogonal frame which are perpendicular
to the direction of the point nodes at $\bm{p}=s\bm{p}_0\equiv s p_{\rm F}\hat{\bm \ell}$, where $s=\pm 1$ is the chirality of the Weyl points, and
$\hat{\bm \ell}=\hat{\bm m}\times\hat{\bm n}$ is the $\ell$-vector. 
$\Delta$ is the superconducting gap, and $p_{\rm F}$ is the Fermi momentum.
Then, the effective Lagrangian for Bogoliubov quasiparticles around $\bm{p}=s \bm{p}_0$ is given by
\begin{eqnarray}
\mathcal{L}_{s}=\bm{p}\cdot\dot{\bm{r}}-\mathcal{H}_{s}(\bm{p},\bm{r}),
\end{eqnarray}
with the Weyl-Bogoliubov Hamiltonian
\begin{eqnarray}
\mathcal{H}_{s}(\bm{p},\bm{r})=s e^{\mu}_aV^a_b\tau^b(p_{\mu} - sp_{0\mu}),
\label{eq:ham1}
\end{eqnarray}
where $V^a_b=\mbox{diag}[\frac{\Delta}{p_F},\frac{\Delta}{p_F},v]$ with $v$ the Fermi velocity,
$\tau^a$ is the Pauli matrix in the particle-hole space, 
and the vielbein $e^{\mu}_a$ is given by 
\beq
(e^{\mu}_{\bar{1}},e^{\mu}_{\bar{2}},e^{\mu}_{\bar{3}})=(m^{\mu},n^{\mu},\ell^{\mu}). 
\label{eq:v}
\eeq
We use greek letter indices $\mu=1,2,3$ as space indices for the laboratory frame, and roman letters $a=\bar{1},\bar{2},\bar{3}$ as indices for a local orthgonal frame. 
The Weyl Hamiltonian must obey the particle-hole symmetry
\beq
\mathcal{C}\mathcal{H}_s({\bm p},{\bm r})\mathcal{C}^{-1}=-\mathcal{H}_{-s}(-{\bm p},{\bm r}),
\label{eq:PHSw}
\eeq
where $\mathcal{C}=K\tau _1$ is the particle-hole conversion operator with $\tau _{\mu}$ the Pauli matrices in the particle-hole space. The particle-hole symmetry guarantees that the Weyl point appears as a pair of ${\bm p}_0$ and $-{\bm p}_0$ and the pairwise Weyl points have opposite chirality, $s=\pm 1$. 

{For $^3$He-A with $\bm{p}_0 = p_{\rm F}\hat{\bm \ell}$, the space-time modulation of the $\hat{\ell}$-texture generates the emergent electromagnetic field, ${\bm B}^{\rm em} = p_{\rm F}{\rm curl}\hat{\bm \ell}$ and ${\bm E}^{\rm em} = p_{\rm F}\partial _t \hat{\bm \ell}$, which directly act on the chirality of Weyl-Bogoliubov quasiparticles. Volovik and his collaborators\cite{volovik81,volovik84,volovik85,volovik86,volovik86-2,balatskii87,volovik95} found the chiral anomaly that the emergent field is responsible for the production of net quasiparticle momentum, $\partial _t {\bm P}_{\rm QP} = \frac{1}{2\pi^2}\int d^3{r}p_{\rm F}\hat{\bm \ell}{\bm E}^{\rm em}\cdot{\bm B}^{\rm em}$. As the quasiparticle momentum ${\bm P}_{\rm QP}$ is equivalent to the quasiparticle mass current in $^3$He-A, the violation of the quasiparticle momentum conservation is compensated by the extra mass current carried by the superfluid vacuum, that is, the anomalous current in Eq.~\eqref{eq:currentMM}, ${\bm j}^{\rm an}= - \frac{\hbar}{2M}C_0\hat{\bm \ell}(\hat{\bm \ell}\cdot{\rm curl}\hat{\bm \ell})$. Below, we show the another contribution of Weyl-Bogoliubov quasiparticles to Eq.~\eqref{eq:currentMM} that the nontrivial torsion field due to the modulation of the vielbein, $e^{\mu}_a$, is accompanied by the extra mass current along ${\rm curl}\hat{\bm \ell}$. This is the chiral magnetic effect due to the torsion field, which can be significant even if $\hat{\bm \ell}\cdot{\rm curl}\hat{\bm \ell}=0$.}


Following the method developed in ref.~\onlinecite{CK}, we obtain the effective Lagrangian for the upper band:
\begin{eqnarray}
\mathcal{L}_{s}=p_{\mu}\dot{r}^{\mu}+\mathcal{E}_s+\mathcal{A}^{+\mu}_{\bm{p} s}\dot{p}_{\mu}+\mathcal{A}^{+}_{\bm{r}s \mu}\dot{r}^{\mu},
\end{eqnarray}
where the Berry connections are $\mathcal{A}^{+\mu}_{\bm{p}s}=i\langle u_{s+}|\partial_{p_{\mu}}|u_{s+}\rangle$,  and
$\mathcal{A}^{+}_{\bm{r}s\mu}=i\langle u_{s+}|\partial_{r^{\mu}}|u_{s+}\rangle$ with
$\mathcal{H}_s|u_{s+}\rangle=\mathcal{E}_s|u_{s+}\rangle $ and 
$\mathcal{E}_s$ is the single-particle energy of Weyl-Bogoliubov quasiparticles with chirality $s$. 

The Berry connection and the Berry curvature in the momentum space charaterizing Weyl fermions appear
when one projects the state into the one of the two energy bands of $\mathcal{H}_{s}(\bm{p},\bm{r})$. 
This approach is justified for Weyl semimetals, when the Fermi level crosses only one band, and the other band is well separated from the Fermi level. However, in the case of Weyl superconductors and Weyl superfluids, the Fermi level crosses the Weyl point at which the lower band touches the upper band, and thus,
 the projection procedure is not justified.
Nevertheless, we exploit this approach to see qualitatively how the Berry curvature plays the role in the response against torsion fields.

In some cases with a spatially varying structure of the gap function such as a vortex, and the Mermin-Ho texture or the Anderson-Toulouse texture of the $\ell$-vector, nonzero torsion appears. The torsion field is defined by 
\begin{eqnarray}
T^a_{\mu\nu}=\frac{\partial e^a_{\nu}}{\partial{r^{\mu}}} -\frac{\partial e^a_{\mu}}{\partial{r^{\nu}}},
\label{eq:torsion}
\end{eqnarray}
where $e^a_{\mu}$ is the inverse of $e^{\mu}_a$.

In the case with nonzero torsion, the Euler-Lagrange equation for $\bm{r}$ and $\dot{\bm{r}}$ is modified as,\cite{Kleinert}
\begin{eqnarray}
\frac{d}{dt} \left(\frac{\partial \mathcal{L}}{\partial \dot{r}^{\mu}}\right)-\frac{\partial \mathcal{L}}{\partial r^{\mu}}=T^{\nu}_{\mu\lambda}\dot{r}^{\lambda}\frac{\partial \mathcal{L}}{\partial \dot{r}^{\nu}},
\end{eqnarray}
with $T^{\nu}_{\mu\lambda}=e^{\nu}_aT^a_{\mu\lambda}$, while that for $\bm{p}$ and $\dot{\bm{p}}$ is not changed.
Then, we obtain the equation of motion for the Weyl-Bogoliubov quasiparticles:
\begin{align}
\frac{d\bm{r}}{dt}=\frac{\partial \mathcal{E}_s}{\partial \bm{p}}-\hat{\Omega}^{+}_{\bm{p}\bm{r}s}\cdot\frac{d\bm{r}}{dt}
-\frac{d\bm{p}}{dt}\times 
\bm{\Omega}^{+}_{\bm{p}\bm{p}s}+\bm{\Omega}_{t\bm{p}s}^{+},
\label{eq:EOM1}
\end{align}
\begin{align}
\frac{d\bm{p}}{dt}=&-\frac{\partial \mathcal{E}_s}{\partial \bm{r}}+\hat{\Omega}^{+}_{\bm{r}\bm{p}s}\cdot\frac{d\bm{p}}{dt} +\frac{d\bm{r}}{dt}\times \bm{\Omega}^{+}_{\bm{r}\bm{r}s}-\bm{\Omega}_{t\bm{r}s}^{+} \nonumber \\
&+\frac{d\bm{r}}{dt}\times \bm{T}^{\mu}(p_{\mu}+\mathcal{A}^{+}_{\bm{r}s\mu}),
\label{eq:EOM2}
\end{align}
where 
the Berry curvatures are,
\begin{eqnarray}
\bm{\Omega}_{\bm{X}\bm{X}s}^{+}=i\langle \nabla_{\bm{X}} u_{s+}| \times | \nabla_{\bm{X}} u_{s+}\rangle,
\end{eqnarray}
\begin{eqnarray}
\bm{\Omega}_{t\bm{X}s}^{+}=i(\langle \partial_t u_{s+}|\nabla_{\bm{X}} u_{s+}\rangle-\langle \nabla_{\bm{X}} u_{s+}|\partial_t u_{s+}\rangle),
\end{eqnarray}
\begin{eqnarray}
(\hat{\Omega}^{+}_{\bm{p}\bm{r}s})_{\alpha\beta}=i(\langle \partial_{p_{\alpha}} u_{s+}| \partial_{r_{\beta}} u_{s+} \rangle -
\langle \partial_{r_{\beta}} u_{s+}| \partial_{p_{\alpha}} u_{s+} \rangle),
\end{eqnarray}
\begin{eqnarray}
\hat{\Omega}^{+}_{\bm{r}\bm{p}s}=-(\hat{\Omega}^{+}_{\bm{p}\bm{r}s})^t,
\end{eqnarray}
with $\bm{X}=\bm{r}, \bm{p}$,
and $(\bm{T}^{\mu})^{\nu}=\frac{1}{2}\epsilon^{\nu\lambda\rho}T^{\mu}_{\lambda\rho}$.
In Eqs.~\eqref{eq:EOM1} and \eqref{eq:EOM2}, all components of vectors are expressed in the laboratory frame. 
We can obtain similar equations also for the lower band. The Berry curvature is $\bm{\Omega}_{\bm{X}\bm{Y}s}^{-}=-\bm{\Omega}_{\bm{X}\bm{Y}s}^{+}$
It is noted that, as seen from (\ref{eq:EOM2}), the torsion generates an effective magnetic field acting on quasiparticles. We denote it as 
\begin{eqnarray}
\bm{\mathcal{B}}=\bm{T}^{\mu}(p_{\mu}+\mathcal{A}^{+}_{\bm{r}\mu})=
\bm{T}^{a}(p_{a}+\mathcal{A}^{+}_{\bm{r}a}). 
\label{eq:torsion-mag}
\end{eqnarray} 
In the following, we consider static inhomogeneity of the order parameter, and neglect  $\bm{\Omega}_{t\bm{p}s}^{+}$ and $\bm{\Omega}_{t\bm{r}s}^{+}$.
Then, as follows from Eqs.~\eqref{eq:EOM1} and \eqref{eq:EOM2},
\begin{align}
\frac{d \bm{r}}{dt}=&
\frac{\partial \mathcal{E}_s}{\partial \bm{p}}
+\frac{\partial \mathcal{E}_s}{\partial \bm{r}}\times \bm{\Omega}^{+}_{\bm{p}\bm{p}s} -\left(\frac{\partial \mathcal{E}_s}{\partial \bm{p}}\cdot\bm{\Omega}^{+}_{\bm{p}\bm{p}s}\right)\tilde{\bm{\mathcal{B}}} \nonumber \\
&
-\hat{\Omega}^{+}_{\bm{p}\bm{r}s}\cdot\frac{d\bm{r}}{dt}
+(\bm{\Omega}^{+}_{\bm{p}\bm{p}s} \cdot \tilde{\bm{\mathcal{B}}})\frac{d \bm{r}}{dt},
\label{eq:dr}
\end{align}
\begin{align}
\frac{d\bm{p}}{dt}=&-\frac{\partial \mathcal{E}_s}{\partial \bm{r}}
+\frac{d\bm{r}}{dt}\times \tilde{\bm{\mathcal{B}}}+\left(\frac{\partial \mathcal{E}_s}{\partial \bm{r}}\cdot \tilde{\bm{\mathcal{B}}}\right)\bm{\Omega}^{+}_{\bm{p}\bm{p}s} \nonumber \\
&
+\hat{\Omega}^{+}_{\bm{r}\bm{p}s}\cdot\frac{d\bm{p}}{dt} 
+(\bm{\Omega}^{+}_{\bm{p}\bm{p}s} \cdot \tilde{\bm{\mathcal{B}}})\frac{d \bm{p}}{dt},
\label{eq:dp}
\end{align}
where $\bm{v}_{\bm{p}s}=\partial \mathcal{E}_s/\partial \bm{p}$, and $\tilde{\bm{\mathcal{B}}}=\bm{\Omega}^{+}_{\bm{r}\bm{r}}+ \bm{\mathcal{B}}$.
In Eq.~\eqref{eq:dr}, the third term of the right-hand side represents the chiral magnetic effect, and the contribution proportional to $\bm{\mathcal{B}}$
corresponds to the torsional chiral magnetic effect found in ref.~\onlinecite{SF}. 
The third term of the right-hand side of Eq.~\eqref{eq:dp} is associated with chiral anomaly; i.e.
the momentum is generated or annihilated at the Weyl points when both an effective magnetic field and a bias potential are applied in the same direction.
It has recently been found that from Eqs.~\eqref{eq:dr} and \eqref{eq:dp} the torsional chiral magnetic effect arising from a ${\rm U}(1)$ phase vortex or lattice strain leads to the negative thermal magnetoresistivity, that is, the anomalous enhancement of longitudinal thermal conductivity along the vortex line.~\cite{kobayashi18}

\subsection{Torsional chiral magnetic effect due to $\ell$-textures}

The current density induced by the torsional field is written as~\cite{CK}
\beq
{j}_{\mu}^{\rm TCME}=\sum _{s,\pm} \int \frac{d^3k}{(2\pi)^3}\left(
{\bm v}^{\pm}_{{\bm k}s}\cdot{\bm \Omega}^{\pm}_{{\bm k}{\bm k}s}
\right)f(\mathcal{E} _{s}) \mathcal{B}^{\mu},
\label{eq:currentTCME}
\eeq
where $f(\epsilon)=1/(e^{\epsilon/T}+1)$ is the Fermi distribution function. Equation~\eqref{eq:currentTCME} reproduces the current expression based on the linear response of the effective action for Weyl-Bogoliubov quasiparticles with respect to the torsional magnetic field.\cite{SF} The effective Hamiltonian in Eq.~\eqref{eq:ham1} possesses anisotropic dispersion of Weyl-Bogoliubov quasiparticles and the Berry curvature is 
\beq
{\bm \Omega}^{\pm}_{{\bm k}{\bm k}s} = \pm s \left( \frac{1}{p_{\rm F}\xi }\right)^2 
\frac{{\bm k}-s{\bm k}_0}{2|\tilde{\bm k}-s\tilde{\bm k}_0|^3},
\eeq
where we have introduced $\tilde{k}_b = V^a_bk_a/v_{\rm F}$ and $\xi = v_{\rm F}/\Delta _{\rm A}$. We also introduce a momentum cutoff $|\tilde{\bm k}-s\tilde{\bm k}_0| < b\Delta _{\rm A}/v_{\rm F}\equiv \Lambda$ for Weyl-Bogoliubov quasiparticles with the chirality $s$, where $b\sim 1$.
A nonzero torsional magnetic field leads to the equilibrium current
\beq
\bm{j}^{\rm TCME}= \frac{v_{\rm F}p_{\rm F}\Lambda}{2\pi^2}{\bm T}^{\bar{3}},
\label{eq:jTCME}
\eeq
where we utilize the particle-hole symmetry in Eq.~\eqref{eq:PHSw}.

As seen from Eqs.~\eqref{eq:torsion} and \eqref{eq:torsion-mag},
in the A-phase of $^3$He, a torsional magnetic field is generated by the rotation of the Weyl points, $\bm{p}_0=p_F\hat{\bm{\ell}}$, as 
\beq
{\bm T}^{\bar{3}}={\rm curl}\hat{\bm \ell}.
\eeq
The real space texture of the $\ell$ vector field is thermodynamically stable in superfluid $^3$He-A under rotation. This is a consequence of the broken gauge-orbit symmetry in the superfluid vacuum;  the ${\rm U}(1)$ phase rotation in Eq.~\eqref{eq:abm} is equivalent to the rotation of the orbital part $\hat{\bm m}+i\hat{\bm n}$ about $\hat{\bm \ell}$. Therefore, the supercurrent can be generated by a variation of $(\hat{\bm m},\hat{\bm n},\hat{\bm \ell})$ without a ${\rm U}(1)$ phase singularity. The $\ell$-texture spontaneously emerges in $^3$He-A under rotation, and continuous skyrmion-like textures provide an elementary building-block for a variety of coreless vortices with the spatially uniform superfluid density.\cite{salomaa,thuneberg} Here we consider skyrmion-vortices with the {N\'{e}el-type} and {Bloch-type} $\hat{\bm \ell}$ textures as in Fig.~\ref{fig:skyrmion}. 

Let us now clarify the torsional field induced by skyrmion vortices. It is convenient to express the orbital part of the order parameter tensor in Eq.~\eqref{eq:abm} in terms of Euler angles $\alpha,\beta,\gamma$ as $\hat{\bm m}+i\hat{\bm n}=e^{-i\gamma}(\hat{\bm m}^{\prime}+i\hat{\bm n}^{\prime})$, where
\begin{gather}
\hat{\bm m}^{\prime}=\cos\beta\cos\alpha\hat{\bm x} + \cos\beta\sin\alpha\hat{\bm y}-\sin\beta\hat{\bm z}, 
\label{eq:m} \\
\hat{\bm n}^{\prime}=-\sin\alpha\hat{\bm x} + \cos\alpha\hat{\bm y}, 
\label{eq:n} 
\end{gather}
and $\hat{\bm \ell} = \cos\alpha\sin\beta \hat{\bm x} + \sin\alpha\sin\beta \hat{\bm y} + \cos\beta\hat{\bm z}$.
The texture is assumed to be translationally invariant along the $z$ axis and be axially symmetric.
The axisymmetric skyrmion texture requires the Euler angles to obey $\alpha\equiv \theta + \alpha_0(r)$, $\gamma =-n\theta$ ($n\in\mathbb{Z}$), and $\beta\equiv \beta(r)$, where $r$ is the distance from the center of the vortex and $\theta$ is the azimuthal angle. For the real-space $\ell$-vector field with axial symmetry, therefore, the parametrization reduces to
\begin{align}
\hat{\bm{\ell}} = \sin\beta\cos\alpha_0 \hat{\bm r} 
+ \sin\beta \sin\alpha_0 \hat{\bm \theta} 
+ \cos\beta \hat{\bm z},
\label{eq:ell}
\end{align}
where $\hat{\bm{\ell}}\equiv \hat{\bm{\ell}}(r)$, $\beta\equiv\beta(r)$, and $\alpha _0\equiv \alpha _0(r)$.
The bending angle is a monotonic function on $r$ which obeys $\beta (r) = 0$ at $r=0$ and $\beta (r) = \pi/2$ at $r=R$, where $R$ determines the size of the skyrmion texture. In Fig.~\ref{fig:skyrmion}, we present the texture of $(\hat{\bm m},\hat{\bm n},\hat{\bm \ell})$ in the skyrmion-vortex with $n=1$: (a) the {N\'{e}el-type} skyrmion with $\alpha = 0$ and (b) the {Bloch-type} skyrmion with $\alpha = \pi /2$. 
From Eqs.~\eqref{eq:ell} and \eqref{eq:v},
the nonzero torsion field in Eq.~\eqref{eq:torsion} can be generated by a skyrmion $\ell$ texture of continuous vortices as 
\begin{align}
{\bm T}^{\bar{3}} &= \beta^{\prime}\sin\beta\hat{\bm \theta} \nn \\
&+ \left[
\left(\beta^{\prime}\cos\beta + \frac{1}{r}\sin\beta \right)\sin\alpha
+ \frac{\alpha^{\prime}}{r}\sin\beta\cos\alpha
\right]\hat{\bm z},
\end{align}
where $\alpha^{\prime}\equiv\partial \alpha /\partial r$ and $\beta^{\prime}\equiv\partial \beta /\partial r$. As shown in Fig.~\ref{fig:skyrmion}, the {N\'{e}el-type} skyrmion texture for $\alpha = 0$ generates the toroidal torsional magnetic field in the $xy$ plane, while the {Bloch-type} skyrmion for $\alpha\neq 0$ is accompanied by the nonzero torsion field along the $z$ axis.

Using the particle density, $\rho = p^3_{\rm F}/3\pi^2$, the TCME is recast into
\begin{eqnarray}
\bm{j}^{\rm TCME}=\left(\frac{b^{\prime}}{p_{\rm F}\xi }\right)\frac{\hbar}{4M}\rho{\rm curl} \hat{\bm \ell},
\label{eq:TCME2}
\end{eqnarray}
where $b^{\prime}\sim 1$ is the dimensionless quantity associated with the cutoff of the Weyl cone. 
%
The skyrmion texture with Eq.~\eqref{eq:ell} therefore gives rise to the in-plane circulating current and the out-of-plane current, depending on $\alpha$. 
We note that $\bm{j}^{\rm TCME}$ in Eq.~\eqref{eq:TCME2} corresponds to the Weyl-Bogoliubov quasiparticle contributions to the second term of the right-hand side of the equilibrium current density in Eq.~\eqref{eq:currentMM}. This is distinct from the chiral anomaly effect due to the emergent electromagnetic fields considered in previous studies.\cite{volovik81,volovik86} In Sec.~\ref{sec:current}, we will discuss the temperature dependence of the equilibrium current density based on the full quantum mechanical calculation.


\section{Chiral fermions in skyrmion-vortex}
\label{sec:chiral}

In this section, we show the emergence of chiral fermions in superfluid $^3$He-A with skyrmion $\ell$-textures. Beyond the semiclassical effective theory, we here utilize the BdG equation which is the fully quantum-mechanical equation for Bogoliubov quasiparticles in superfluid $^3$He-A. 


We start with the Hamiltonian for the equal spin pairing state, $A_{\mu i}({\bm r})=\Delta _i({\bm r})\hat{d}_{\mu}$,
\begin{align}
\mathcal{H} &= \int d {\bm r} \psi^{\dag}({\bm r}) \varepsilon (-i{\bm \nabla})\psi ({\bm r}) \nn \\
&+
\frac{1}{2} \int d{\bm r}\left[
\psi^{\dag}({\bm r})\left\{\Delta _j({\bm r}), \frac{i}{p_{\rm F}}\partial _j\right\} \psi^{\dag}({\bm r})
+ {\rm h.c.}\right],
\label{eq:h0}
\end{align}
where $\psi _a$ and $\psi^{\dag}_a$ denote the fermionic field operators. As $^3$He-A is the equal spin pairing state, we omit the spin degrees of freedom. The single-particle Hamiltonian density represents fermions with mass $M$, 
$\varepsilon (-i{\bm \nabla}) = -{\bm \nabla}^2/2M - \mu$. 
The repeated Roman and Greek indices imply the sum over the spin degrees of freedom and $(x, y, z)$, respectively. 

Now, let us introduce the Bogoliubov transformation of the fermion operator ${\bm \Psi} \equiv [\psi,\psi^{\dag}]^{\rm T}$, 
\beq
{\bm \Psi}({\bm r}) 
= \sum _{E>0} \left[ {\bm \varphi}_{E} ({\bm r}) {\eta}_E + \mathcal{C}{\bm \varphi}_{E} ({\bm r}) {\eta}^{\dag}_E\right],
\label{eq:bogo}
\eeq
where ${\eta}_E$ and $\eta^{\dag}_E$ stand for Bogoliubov quasiparticle operators with the energy $E$ that satisfy fermionic anticommutation relations. We notice that Eq.~\eqref{eq:bogo} obeys the particle-hole symmetry, $\mathcal{C}{\bm \Psi} = {\bm \Psi}$, with $\mathcal{C}=\tau _x K$ where $K$ is the complex conjugation operator. Substituting Eq.~\eqref{eq:bogo} into the Hamiltonian \eqref{eq:h0}, one obtains the Bogoliubov-de Gennes (BdG) equations 
\begin{align}
\mathcal{H}_{\rm BdG}({\bm r}){\bm \varphi}_{E} ({\bm r}) = E {\bm \varphi}_{E} ({\bm r}),
\label{eq:bdg1}
\end{align}
with 
\begin{align}
\mathcal{H}_{\rm BdG}({\bm r})=
\left(
\begin{array}{cc}
\varepsilon (-i{\bm \nabla}) & \frac{1}{2}\sum _j \left\{\Delta _j, \frac{i}{p_{\rm F}}\partial _j\right\} \\
\frac{1}{2}\sum _j \left\{\Delta^{\ast}_{j}, \frac{i}{p_{\rm F}}\partial _j\right\} & -\varepsilon(-i{\bm \nabla}) 
\end{array}
\right),
\label{eq:bdg}
\end{align}
{where $\{,\}$ denotes an anti-commutator.} To make the Bogoliubov transformation canonical, the quasiparticle wavefunction must satisfy the orthonormal conditions
$
\int d{\bm r} {\bm \varphi}^{\dag}_{E} ({\bm r}) {\bm \varphi}_{E^{\prime}} ({\bm r}) = \delta _{E,E^{\prime}}
$.


\subsection{Symmetry of Weyl-Bogoliubov quasiparticles in skyrmion-vortices}

To clarify the symmetry of Bogoliubov quasiparticles in the presence of a skyrmion $\ell$-texture, we start with the symmetry group relevant to the classification of axisymmetric vortices in superfluid $^3$He,
\beq
G_{\rm v} = D_{\infty,h} \times t^z \times T \times {\rm U}(1)_{\varphi},
\eeq
where $D_{\infty,h}$ contains the group of rotations about the vortex line ($\hat{\bm z}$), rotations about an axis perpendicular to $\hat{\bm z}$, and space inversion, and $t^z$ represents the translational symmetry of the skyrmion-vortex along $\hat{\bm z}$. The time reversal symmetry, $T$, transforms the order parameter tensor as $A_{\mu i}\mapsto A^{\ast}_{\mu i}$. The generator of the continuous rotation symmetry about $\hat{\bm z}$ is expressed as $e^{i\hat{Q}\varphi}$, where $\hat{Q}\equiv \hat{L}_z - n \hat{I}$ ($n\in\mathbb{Z}$) is the combination of the orbital angular momentum operator $\hat{L}_z$ and the ${\rm U}(1)$ phase rotation operator. 
The order parameter for axisymmetric vortices satisfies $\hat{Q}\Delta _j({\bm r})=0$. For the skyrmion-vortex states in Fig.~\ref{fig:skyrmion}, the orbital components of the order parameter are given by 
\beq
{\bm \Delta}({\bm r}) = \Delta _{\rm A}(T) e^{i\theta} [\hat{\bm m}^{\prime}(r)+i\hat{\bm n}^{\prime}(r)],
\label{eq:delta}
\eeq
with Eqs.~\eqref{eq:m} and \eqref{eq:n}. In this paper, we focus on the $n=1$ case. 

It is convenient to transform Eq.~\eqref{eq:delta} into the eigenstates of the orbital angular momentum $l=+1, 0, -1$ as $(\Delta _{+1},\Delta _0, \Delta _{-1})$. This transforms the off-diagonal component of Eq.~\eqref{eq:bdg} as $\sum _j \{\Delta _j, \frac{i}{p_{\rm F}}\partial _j\}\mapsto \sum _{l}\{\Delta _{l}({\bm r}),\mathcal{Y}_{1,l}({\bm \partial})\}$, where  $\mathcal{Y}_{L,l}({\bm \partial})$ is the spherical harmonic function of degree $L$ obtained by replacing $\hat{\bm p}\rightarrow -i{\bm \partial}/p_{\rm F}$. The phase factor $e^{il\theta}$, which appears in $\mathcal{Y}_{1,l}$, is compensated by the winding of $(\hat{\bm m}^{\prime},\hat{\bm n}^{\prime})$ in $\Delta _{l}({\bm r})$ and the ${\rm U}(1)$ phase factor $e^{i\theta}$ is factorized from the off-diagonal component in Eq.~\eqref{eq:bdg} at all. 
The quasiparticle wavefunctions are then factorized in terms of the azimuthal quantum number $m\in \mathbb{Z}$ and axial quantum number $k$ as
\beq
{\bm \varphi}_E({\bm r}) = e^{ik z}\left( 
\begin{array}{cc}
u_{nmk} (r) e^{im\theta}\\
v_{mmk} (r)e^{i(m-1)\theta}
\end{array}
\right).
\label{eq:cylinder}
\eeq
Here we impose the periodic boundary condition along the axial direction, ${\bm \varphi}_E(x,y,z) = {\bm \varphi}_E(x,y,z+Z)$,
which implies $k = 2\pi n_z/Z$ with $n_z\in \mathbb{Z}$. By using the factorization in Eq.~\eqref{eq:cylinder}, Eq.~\eqref{eq:bdg} is reduced to the one-dimensional differential equation for $[u_{nmk} (r), v_{nmk} (r)]$, where the set of the quantum numbers is given as $(n,m,k)$. 
The differential equation in the cylindrical coordinate is solved by expanding the wavefunctions with the orthonormal basis. The set of the basis functions is constructed with the Bessel function and the $i$-th zeros of the $\nu$ Bessel function, $J_{\nu}(r)$ and $\alpha^{\nu}_i$, as $\{C^{\nu}_j J_{\nu}(\alpha _j^{\nu}r/R)\}_{j=1,\cdots,N}$, {where $C^{\nu}_j$ and $N$ denote the normalization constant and the number of the orthonormal functions, respectively. This imposes the rigid wall boundary condition,} $u_{nmk} (r=R)=v_{nmk} (r=R)=0$. The Bessel function expansion then reduces Eq.~\eqref{eq:bdg} to the $2N\times 2N$ eigenvalue equation.~\cite{mizushimaPRA10,matsumoto,gygi} {We take $R=50$-$200k^{-1}_{\rm F}$ and $N=600$.}

\begin{table}[t!]
\caption{\label{table1}Classification of skyrmion-vortex textures in terms of the discrete symmetries. The ``N\'{e}el'', ``Bloch'', and ``twist'' textures correspond to $\alpha=0$, $\alpha=\pi/2$, and $\alpha(r)$, respectively, $mathcal{B}$ is the torsional magnetic field due to the $\ell$-texture, and ``ZES'' denotes the distribution of the zero energy states. 
}
\begin{ruledtabular}
\begin{tabular}{ccccccccc}
& class & $\ell$-texture & $P_1$ & $P_2$ & $P_3$ & $\mathcal{B}\parallel \hat{\bm z}$ & ZES & \\
\hline
& $v$ & N\'{e}el & -- & $\surd$ & -- & -- & flat & \\
& $w$ & Bloch & -- & -- & $\surd$ & $\surd$ & point & \\
& $uvw$ & twist & -- & -- & -- & $\surd$ & point & \\
\end{tabular}
\end{ruledtabular}
\end{table}

Apart from the continuous symmetry, there are discrete symmetries, which leave axisymmetric vortex order parameter invariant. First we note that the BdG Hamiltonian in Eq.~\eqref{eq:bdg} always satisfies the particle-hole symmetry
\beq
\mathcal{C}\mathcal{H}_{\rm BdG}({\bm r})\mathcal{C}^{-1} = - \mathcal{H}_{\rm BdG}({\bm r}).
\eeq
This results in relation \eqref{eq:PHSw}. The symmetry guarantees that the eigenstate of the BdG equation must appear as a pair of the positive and negative energy states. The positive energy state with $E_{n}(m,k)$ and ${\bm \varphi}_{n,m,k}(r)\equiv[u_{nmk} (r),v_{nmk} (r)]^{\rm T}$ has the particle-hole symmetric partner with $-E_{n}(-m+1,-k)$ and $\mathcal{C}{\bm \varphi}_{n,-m+1,-k}(r)$. 

The other discrete symmetries relevant to the vortex classification are given by three operators, $\{P_1,P_2,P_3\}$.~\cite{salomaaPRB85,salomaaPRL83,salomaa,mizushimaJPSJ16} $P_1$ is the space inversion operator. The order-2 antiunitary operator, $P_3$, is the combination of the time reversal symmetry and $\pi$-rotation about any axis perpendicular to the vortex line. The $P_2$ symmetry is defined as $P_2=P_1P_3$, which is the combination of the time-reversal symmetry and mirror reflection symmetry in a plane that contains the vortex line. Axisymmetric continuous vortices with skyrmion-like $\ell$-texture are classified in terms of these discrete symmetries into three categories: $v$-, $w$-, and $uvw$-vortices.~\cite{salomaa} The order parameters of the $v$- and $w$-vortices are invariant under the $P_2$ and $P_3$ symmetry, respectively, while the $uvw$-vortex class spontaneously breaks all the discrete symmetries. 

Let us define the operator $M=i\sigma _y$ that denotes the mirror reflection in the $xz$ plane. The operator flips the spin, momentum, and spatial coordinate as ${\bm \sigma}\mapsto (-\sigma _x,\sigma _y,-\sigma _z)$, ${\bm k}\mapsto (k_x,-k_y,k_z)$, and ${\bm r}\mapsto (x,-y,z)$, respectively. The $P_2$ operator is defined as the combination of $M$ and the time-reversal operator $\mathcal{T}=-i\sigma _yK$. For $\hat{\bm d}\parallel\hat{\bm z}$, the $P_2$ operator flips the triad as $\hat{\bm m} (r,\theta,z)\mapsto [\hat{m}_x(r,-\theta,z),-\hat{m}_y(r,-\theta,z),\hat{m}_z(r,-\theta,z)]$, $\hat{\bm n} (r,\theta,z)\mapsto [-\hat{n}_x(r,-\theta,z),\hat{n}_y(r,-\theta,z),-\hat{n}_z(r,-\theta,z)]$, and $\hat{\bm \ell} (r,\theta,z)\mapsto [\hat{\ell}_x(r,-\theta,z),-\hat{\ell}_y(r,-\theta,z),\hat{\ell}_z(r,-\theta,z)]$. Similarly, the $P_3$ symmetry relates the triad at $(r,\theta,z)$ to $(\hat{m}_x,-\hat{m}_y,\hat{m}_z)$, $(-\hat{n}_x,\hat{n}_y,-\hat{n}_z)$, and $(\hat{\ell}_x,-\hat{\ell}_y,\hat{\ell}_z)$ at $(r,\pi-\theta,-z)$. 
The {N\'{e}el-type} skyrmion-vortex ($v$-vortex) with $\alpha=0$ in Fig.~\ref{fig:skyrmion}(a) spontaneously breaks the $P_1$ symmetry but maintains the $P_2$ symmetry. The vortex state with the {Bloch-type} skyrmion texture ($\alpha=\pi/2$) in Fig.~\ref{fig:skyrmion}(b) belongs to the $w$-vortex class with $P_3$ symmetry. The $uvw$-vortex class can be realized by twisting the skyrmion $\ell$-texture so as to satisfy the conditions $\alpha(0) =\pi/2$ and $\alpha(R)=0$. In Table~\ref{table1}, we summarize the possible classification of skrymion-vortices in terms of the discrete symmetries.

The $P_2$ symmetry imposes an important constraint on the energy spectrum of the Bogoliubov quasiparticles so as to prohibit the equilibrium current along the axial direction. Axisymmetric $v$-vortices must satisfy the relation 
\beq
E_{n}(m,k) = E_{n}(m,-k) = - E_{n}(-m+1,-k).
\label{eq:p2}
\eeq
The first equality results from the $P_2$ symmetry, while the second equality reflects the particle hole symmetry. Hence, the Bogoliubov quasiparticle spectrum in the {N\'{e}el-type} skyrmion-vortex with $\alpha=0$ is an even function on $k$ and the current flow along $\hat{\bm z}$ is prohibited. In contrast, as the $P_3$ symmetry does not impose any constraints on the eigenvalues, the {Bloch-type} and twisted skyrmion $\ell$-textures may generate the equilibrium current along the axial direction. 
%

\begin{figure}[t]
\includegraphics[width=85mm]{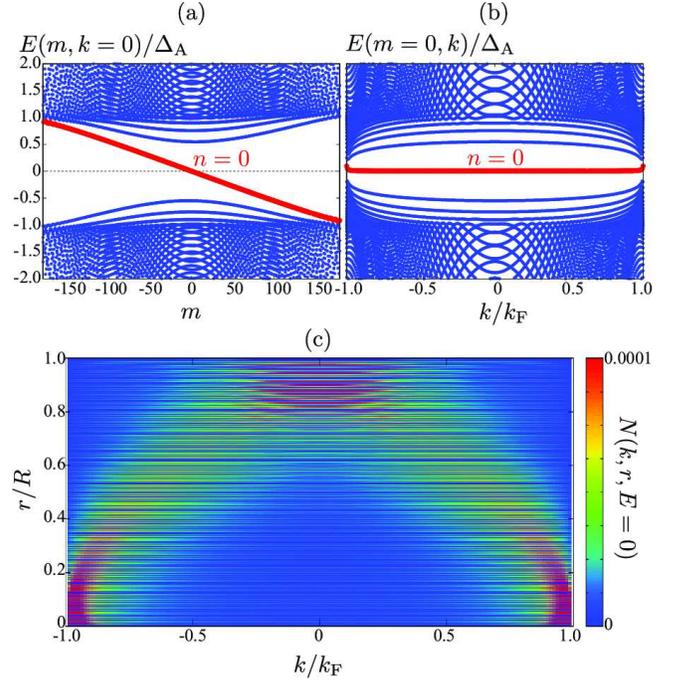}
\caption{Bogoliubov quasiparticle spectra for {N\'{e}el-type} skyrmion ($\alpha=0$) at $E_n(m,k=0)$ (a) and $E_n(m=0,k)$ (b).
(c) $k$-resolved zero-energy local density of states, ${N}(k,r,E=0)$ representing the spatial distribution of the chiral fermion.
We fix $R=200k^{-1}_{\rm F}$ and $p_{\rm F}\xi = 20$.}
\label{fig:chiral}
\end{figure}

\subsection{Chiral fermions and real space texture of Weyl points}
\label{sec:spectra}

Let us consider the superfluid $^3$He-A confined in a cylinder with radius $R$. The triad $(\hat{\bm m},\hat{\bm n},\hat{\bm \ell})$ parameterized as Eqs.~\eqref{eq:m}, \eqref{eq:n}, and \eqref{eq:ell} slowly varies from $\hat{\bm \ell}=\hat{\bm z}$ at $r=0$ to $\hat{\bm \ell}=\hat{\bm r}$ at $r=R$. Hence, the bending angle is given by
$\beta(r) = \frac{\pi}{2R}r$. We here set the angle as $\alpha (r)=0$ for the {N\'{e}el-type} skyrmion ($v$-vortex) and $\alpha (r)= \frac{\pi}{2}(1-r/R)$ for the twisted skyrmion ($uvw$-vortex). The size of the half-skyrmion is set to be larger than the superfluid coherence length $\xi$, $R\gtrsim 10\xi $ where $\xi \equiv v_{\rm F}/\Delta _{\rm A} > k^{-1}_{\rm F}$.

In Fig.~\ref{fig:chiral}, we show the Bogoliubov quasiparticle spectra obtained by diagonalizing Eq.~\eqref{eq:bdg} with the {N\'{e}el-type} skyrmion textures ($v$-vortex). The Bogoliubov spectrum is asymmetric with respect to the azimuthal quantum number $m$ and lowest branch ($n=0$) crosses the zero energy. As mentioned in Sec.~\ref{sec:TCME}, the Weyl-Bogoliubov quasiparticles around point nodes experience the torsional magnetic field ${\bm T}^{\bar{3}}= {\bm \nabla}\times {\bm \ell}$. 
For the {N\'{e}el-type} $\ell$ skyrmion with $\alpha = 0$, the toroidal torsional magnetic field, ${\bm T}^{\bar{3}}\propto\hat{\bm \theta}$, leads to the emergence of the Landau levels linearly dispersing from $m=0$. The lowest energy branch crossing the Fermi level in Fig.~\ref{fig:chiral}(a) is asymmetric with respect to $m$
\beq
E_0(m,k) = - v\left( m - \frac{1}{2}\right),
\label{eq:chiral}
\eeq
which is identified as the chiral fermion due to the emergent toroidal field. 
The group velocity, $-v<0$, is an order of $\Delta _{\rm A}/k_{\rm F}$. Figure~\ref{fig:chiral}(b) shows the dispersion of the Bogoliubov spectrum with respect to the axial momentum $k$ at $m=0$, which satisfies the $P_2$ symmetry constraint in Eq.~\eqref{eq:p2}. The almost flat dispersion of the lowest eigenstates indicates that the chiral branch in Eq.~\eqref{eq:chiral} exists within $|k|< k_{\rm F}$. Hence, the spectral asymmetry in the {N\'{e}el-type} skyrmion-vortex leads to the equilibrium current along the azimuthal direction and the $P_2$ symmetry prohibits the flow along the axial direction. 

To capture the spatial distribution of the chiral fermions, we show in Fig.~\ref{fig:chiral}(c) the $k_z$-resolved zero-energy local density of states,
$N (k_z,r,E)$,~\cite{ichiokaPRB10}
\begin{align}
N(k,r,E) = \sum_{E>0}{}^{\prime}&\left[
|u_{nmk}(r)|^2\delta (E-E_{n}(m,k))\right. \nn \\
& \left. +|v_{nmk}(r)|^2\delta (E+E_{n}(m,k))
\right],
\label{eq:kz}
\end{align}
where $\sum^{\prime}_{E>0}$ stands for the sum over $(n,m)$ that satisfies $E_{n}(m,k)>0$. 
The peak amplitude in the plane $(k,r)$ shifts from $r=R$ at $k=0$ to $r=0$ at $k=\pm k_{\rm F}$. The spectral evolution reflects the spatial profiles of the {N\'{e}el-type} skyrmion $\ell$-texture that smoothly tilts from the axial direction to the {N\'{e}el-type} direction. Therefore, the asymmetric branch crossing $E=0$ in Fig.~\ref{fig:chiral} is attributed to the Weyl-Bogoliubov quasiparticles bound to the Weyl points and the $\ell$-vector texture leads to spatially inhomogeneous structures of Weyl bands in the real coordinate space.

\begin{figure}[t]
\includegraphics[width=85mm]{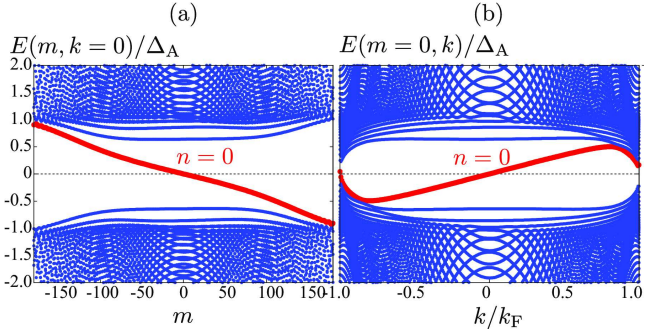}
\caption{Bogoliubov quasiparticle spectra for twisted skyrmion ($\alpha=\pi/2(1-r/R)$) at $E_n(m,k=0)$ (a) and $E_n(m=0,k)$ (b).
We fix $R=200k^{-1}_{\rm F}$ and $p_{\rm F}\xi = 20$.}
\label{fig:weyl}
\end{figure}

Figure \ref{fig:weyl} shows the Bogoliubov quasiparticle spectra for the twisted skyrmion $\ell$-texture. It is seen that the chiral fermion branch with the spectral asymmetry exists in the $k$ direction as well as the azimuthal momentum $m$. In order to satisfy the rigid wall boundary condition, $\hat{\bm \ell}=\hat{\bm r}$, at $r=R$, we set the azimuthal angle as $\alpha(r)=(1-r/R)\pi/2$. The resulting $\ell$ texture generates the torsional magnetic field long the axial direction in addition to the azimuthal direction. This is categorized to the $uvw$-vortex class which holds neither the $P_2$ symmetry nor $P_3$ symmetry. The symmetry relation in Eq.~\eqref{eq:p2} can be violated and the spectral asymmetry along $k$ is responsible for the equilibrium current along the axial direction.

\section{Torsional Chiral Magnetic Effect and mass current in skyrmion-vortex}
\label{sec:current}

In the previous section, we have demonstrated that the chiral fermion branches emerge in the Bogoliubov quasiparticle spectrum under skyrmion-like $\ell$-textures. Using the full quantum mechanical BdG equation, in this section, we show that low-lying Weyl-Bogoliubov quasiparticles dominantly contribute to the current density in the weak coupling regime, while the contributions from continuum states become significant as the topological phase transition is approached. Here we introduce the dimensionless parameter $p_{\rm F}\xi =2 E_{\rm F}/\Delta _{\rm A}$ so as to quantify the quantum corrections to the quasiclassical limit ($p_{\rm F}\xi \gg 1$). 

\subsection{N\`{e}el-type skyrmion}

We define the mass current density ${\bm j}({\bm r})$ as the linear response of the thermodynamic potential with respect to an infinitesimal flow ${\bm v}$, ${j}_{\mu}=(\delta\langle \mathcal{H}\rangle /\delta {v}_{\mu})_{v=0}$, where the Hamiltonian under a homogeneous velocity field is given by a Galilean transformation $-i{\bm \nabla} \mapsto -i{\bm \nabla}-M{\bm v}$. The current density is then given by
$
j_{\mu}({\bm r})= -i \langle
\psi^{\dag}({\bm r}) \partial _{\mu}\psi({\bm r})
- \psi ({\bm r}) \partial _{\mu} \psi^{\dag}({\bm r})
\rangle
$.
In terms of the Bogoliubov quasiparticle wavefunctions ${\bm \varphi}_E=[u_E,v_E]^{\rm T}$, this is rewritten to 
\begin{align}
j_{\mu}({\bm r}) = 2\sum _{E>0} & \left[
{\rm Im}\left\{u^{\ast}_E({\bm r})\partial _{\mu} u_E({\bm r})\right\}f(E) \right. \nn \\
& \left. +{\rm Im}\left\{v_E({\bm r})\partial _{\mu} v^{\ast}_E({\bm r})\right\}f(-E)
\right],
\label{eq:current}
\end{align}
where the factor ``2'' arises from the spin degeneracy in the equal spin pairing state and $f(E) = 1/(e^{E/T}+1)$ is the Fermi distribution function at temperature $T$. Owing to the particle-hole symmetry, the azimuthal current density in Eq.~\eqref{eq:current}, or the angular momentum density $({\bm r}\times {\bm j})_z=rj_{\theta}(r)$, is recast into 
\beq
rj_{\theta}(r) = 
-2\sum _{E>0}
(m-1)|v_E(r) |^2
= 2\sum _{E < 0} m |u_E(r)|^2,
\eeq
at $T=0$, where $n ({\bm r})= 2\langle\psi^{\dag}\psi \rangle$ is the particle density and $\sum _{E>0}$ stands for the sum over $(n,m,k)$ within $E_n(m,k)>0$. 
As shown in Fig.~\ref{fig:chiral}(a), the chiral fermion states with $m \ge 1$ have negative energy and thus are occupied at $T=0$. These chiral fermion states make a positive contribution to the mass current density along the azimuthal direction, $j_{\theta}(r)$. In the {N\'{e}el-type} skyrmion-vortex, therefore, they produce an azimuthal mass current in the same sense as the torsional field, ${\bm T}^{\bar{3}}={\rm curl}\hat{\bm \ell}$.

\begin{figure}[t]
\includegraphics[width=70mm]{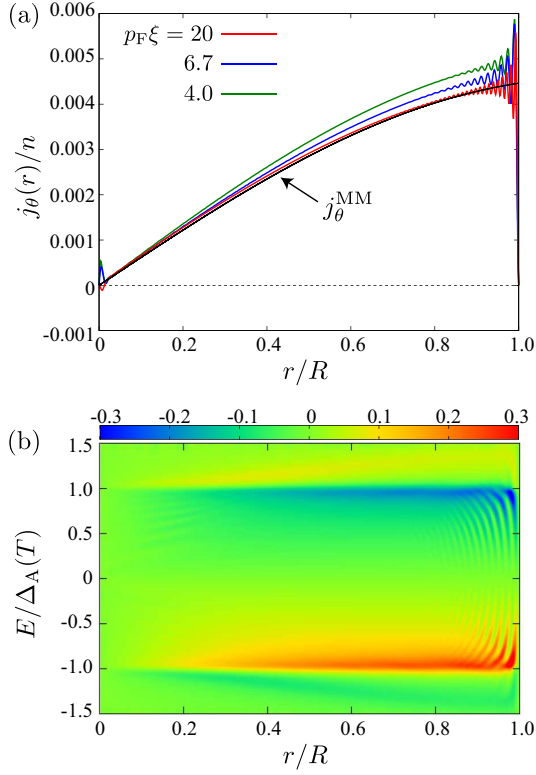}
\caption{(a) Current density in the {N\'{e}el-type} skyrmion-vortex, where $n$ is the particle density. 
(b) $E$-resolved current density profile, $j_{\theta}(r,E)$, for $k_{\rm F}\xi = 20$. Here we fix $T=0$, $R=200k^{-1}_{\rm F}$, and $\mu=E_{\rm F}$.}
\label{fig:radial}
\end{figure}

In Fig.~\ref{fig:radial}(a), we plot the current density, $j_{\theta}(r)$, in the {N\'{e}el-type} skyrmion-vortex ($v$-vortex) at $T=0$ for $p_{\rm F}\xi =20$, $6.7$, and $4.0$. For comparison, we plot the current density obtained from the gradient expansion, ${\bm j}^{\rm MM}$, with $C_0=\rho$. As the {N\'{e}el-type} skyrmion-vortex always satisfies $\hat{\bm \ell}\perp {\rm curl}\hat{\bm \ell}$, this configuration is free from the issue on the anomalous term ${\bm j}^{\rm an}$. It is seen from Fig.~\ref{fig:radial}(a) that the current density obtained from the BdG equation is in good agreement with ${\bm j}^{\rm MM}$ in the weak coupling regime $p_{\rm F}\xi =20$, while $j_{\theta}(r)$ deviates from ${\bm j}^{\rm MM}$ as ${ p}_{\rm F}\xi $ decreases. 
To clarify the Weyl-Bogoliubov quasiparticle contributions, we introduce the $E$-resolved current density as 
\begin{align}
j_{\mu}({\bm r},E) =& 2\sum _{E_i>0}  \left[
{\rm Im}\left\{u^{\ast}_{E_i}({\bm r})\partial _{\mu} u_{E_i}({\bm r})\right\}\delta (E-E_i) \right. \nn \\
& \left. +{\rm Im}\left\{v_{E_i}({\bm r})\partial _{\mu} v^{\ast}_{E_i}({\bm r})\right\}\delta (E+E_i)
\right],
\label{eq:jE}
\end{align}
where $E_i\equiv E_n(m,k)$. The current density is obtained by integrating ${\bm j}(r,E)$ over $E$ as $j_{\mu}({\bm r})=\int dE j_{\mu}({\bm r},E)f(E)$. For numerical calculations, the $\delta$-function in Eq.~\eqref{eq:jE} is replaced by the Lorentzian function with the width $0.025\Delta _{\rm A}$. Figure~\ref{fig:radial}(b) shows the $E$-resolved current density for the {N\'{e}el-type} skyrmion-vortex at $p_{\rm F}\xi = 20$. The mass current density may be decomposed into two contributions, ${\bm j}={\bm j}^{\rm Weyl}+{\bm j}^{\rm cont}$. The contribution arising from Weyl-Bogoliubov quasiparticles, ${\bm j}^{\rm Weyl}$, is defined as 
\beq
{\bm j}^{\rm Weyl}({\bm r}) \equiv \int^{\Delta _{\rm A}}_{-\Delta _{\rm A}} dE {\bm j}({\bm r},E)f(E),
\eeq
and ${\bm j}^{\rm cont}\equiv {\bm j}-{\bm j}^{\rm Weyl}$ is the current carried by continuum states. It is seen from Fig.~\ref{fig:radial}(b) that Weyl-Bogoliubov quasiparticle states, including the chiral branch, dominantly contribute to the mass current. However, the continuum states within $|E|>\Delta _{\rm A}$ make non-vanishing contributions to the mass current. They satisfy ${\bm j}^{\rm cont}({\bm r})\approx - {\bm j}^{\rm Weyl}/2$ for $p_{\rm F}\xi \gg 1$, and lead to the counter flow to the Weyl-Bogoliubov quasiparticle flow.

The backflow of the continuum states is also pointed out in the edge mass current with spatially polarized $\hat{\ell}$-vectors~\cite{saulsPRB11,tsutsumiPRB12} and the surface spin current in the B-phase of the superfluid $^3$He.~\cite{wupRB13,tsutsumiJPSJ12,mizushimaJPCM15} The main contributions to the mass/spin currents originate in chiral/helical fermion states that are the topologically protected Andreev bound states at the edge. As pointed out by Stone and Roy,~\cite{stonePRB04} however, the bound states are not the only contribution. Another contribution results from the continuum states affected by the formation of the Andreev bound states. The contribution cancels the bound states, and the edge current arising from the bound states alone differs from the actual edge current by a factor of $2$. The $E$-resolved current density in Fig.~\ref{fig:radial}(b) resembles to the behavior of the edge mass/spin current, where the continuum states are affected by the existence of the nontrivial $\hat{\ell}$-texture and weakens the flow arising from the chiral fermion states. 

\begin{figure}[t]
\includegraphics[width=80mm]{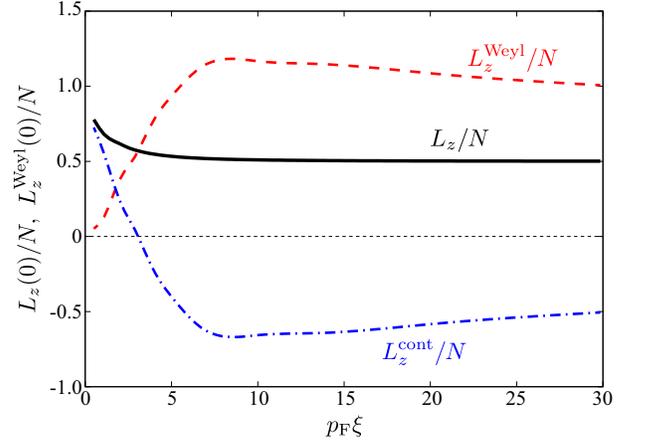}
\caption{$p_{\rm F}\xi$-dependence of the total angular momentum at $T=0$: $L_z/N$ (solid line), $L^{\rm Weyl}_z/N$ (dashed line), and $L_z^{\rm cont}/N$ (dashed-dotted line). $N$ is the total particle number, $N=\int n({\bm r})d{\bm r}$, and $L^{\rm Weyl}_z$ stands for the contributions of low-lying Weyl-Bogoliubov quasiparticles within $|E_n(m,k)|<\Delta _{\rm A}$ to the total angular momentum. }
\label{fig:lz0}
\end{figure}

To capture the contribution of the Weyl-Bogoliubov quasiparticles more systematically, we calculate the total angular momentum per particle at $T=0$. The total angular momentum is defined as 
\beq
L_z = \int ({\bm r}\times {\bm j})_z d{\bm r},
\eeq
and the total particle number is given by $N=\int n ({\bm r})d{\bm r}$. The angular momentum arising from the Weyl-Bogoliubov quasiparticles within $|E|< \Delta _{\rm A}$ is given by $L^{\rm Weyl}_z = \int ({\bm r}\times {\bm j}^{\rm Weyl})_z d{\bm r}$ and the contribution of the continuum states is $L_z^{\rm cont}\equiv L_z-L_z^{\rm Weyl}$.  In Fig.~\ref{fig:lz0}, we plot the $p_{\rm F}\xi$-dependence of the total angular momentum. The total angular momentum in an isolated Mermin-Ho vortex or a {N\'{e}el-type} skyrmion-vortex was also calculated by Nagai~\cite{nagai12,nagaiJLTP14} using the quasiclassical theory, which reproduces the McClure-Takagi prediction,~\cite{MT} $L_z=\hbar N/2$, at $T\rightarrow 0$. In Fig.~\ref{fig:lz0}, the total angular momentum approaches $L_z= \hbar N /2$ at $T=0$ for $p_{\rm F}\xi \gg 1$. The angular momentum associated with the Weyl-Bogoliubov quasiparticles differs from the total angular momentum by a factor of $2$, which implies the existence of backflow arising from the continuum states, $L_z^{\rm cont}=-L_z^{\rm Weyl}/2 = -N\hbar/2$. 

In Fig.~\ref{fig:lz0}, however, the numerical calculation of the BdG equation in the vicinity of the topological phase transition ($p_{\rm F}\xi =0$) shows that $L_z/N$ gradually increases from $\hbar/2$ as $p_{\rm F}\xi$ decreases. Although $L_z/N$ almost stays constant for $p_{\rm F}\xi \gtrsim 5$, we find two characteristic behaviors in $L_z^{\rm Weyl}$ and $L_z^{\rm cont}$; (i) the Weyl-Bogoliubov quasiparticle contribution, $L_z^{\rm Weyl}/N$, exhibits the nonmonotonic behavior as a function of $p_{\rm F}\xi$ and has a maximum around $p_{\rm F}\xi = 10$. (ii) The continuum contribution, $L_z^{\rm cont}/N$, changes its sign around $p_{\rm F}\xi = 3$ and makes a dominant contribution to $L_z$ in $p_{\rm F}\xi \lesssim 1$.  
As for (i), the characteristic $p_{\rm F}\xi$-dependence of $L^{\rm Weyl}_z$ enables one to discriminate the contribution of TCME from other low-lying quasiparticle contributions. 
For $p_{\rm F}\xi \gtrsim 10$, the lower energy part of the angular momentum, $L_z^{\rm Weyl}$, is approximately decomposed into $L_z^{\rm Weyl}\sim \hbar N + L_z^{\rm TCME}$, where $L^{\rm TCME}_z \equiv \int ({\bm r}\times {\bm j}^{\rm TCME})_zd{\bm r}\propto 1/(p_{\rm F}\xi)$ is the angular momentum arising from the torsion-induced current ${\bm j}^{\rm TCME}$ in Eq.~\eqref{eq:jTCME}. The TCME vanishes at the weak coupling limit $p_{\rm F}\xi \gg 1$, while it makes a significant contribution to $L_z^{\rm Weyl}$ as $p_{\rm F}\xi $ decreases. The contribution of ${\bm j}^{\rm TCME}$ is consistent with the increase behavior of $L^{\rm Weyl}_z$ with decreasing $p_{\rm F}\xi $ within $p_{\rm F}\xi \gtrsim 10$. We note that the anomalous enhancement of $L_z^{\rm Weyl}$ is compensated by $L^{\rm cont}_z$ and the resultant $L_z/N$ stays constant at $L_z/N=\hbar/2$ for $p_{\rm F}\xi \gtrsim 10$. 

As Eq.~\eqref{eq:jTCME} is derived from the semiclassical equations of motion for Weyl-Bogoliubov quasiparticles, however, we must be careful about the applicability of Eq.~\eqref{eq:jTCME}. Equation \eqref{eq:jTCME} is applicable only to the large $p_{\rm F}\xi$ regime and fails down in the quantum regime $p_{\rm F}\xi \sim O(1)$. Figure~\ref{fig:lz0} indeed shows that $L_z^{\rm Weyl}$ vanishes at $p_{\rm F}\xi=0$ where the bulk excitation gap closes and topological phase transition occurs.


\begin{figure}[t!]
\includegraphics[width=80mm]{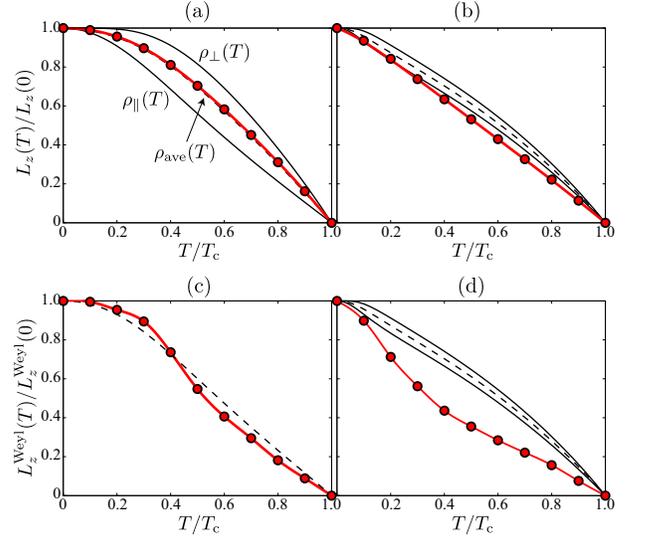}
\caption{(a,b) $T$-dependence of total angular momentum, $L_z(T)$, for the {N\'{e}el-type} skyrmion vortex (red line): (a) $p_{\rm F}\xi = 20$ and (b) $2.0$. (c,d) $T$-dependence of Weyl-Bogoliubov quasiparticle contributions, $L_z^{\rm Weyl}(T)$ (red line), for $p_{\rm F}\xi = 10$ (a) and $2.0$ (b). In all figures, the solid lines are the components of the superfluid density tensor parallel and perpendicular to $\hat{\bm \ell}$, $\rho _{\parallel}$ and $\rho _{\perp}$. The dashed line corresponds to $\rho _{\rm ave}$, where $\rho _{\rm ave}= L_z^{\rm Cross}(T)/L_z^{\rm Cross}(0)$ for $p_{\rm F}\xi \gg 1$. 
}
\label{fig:lz}
\end{figure}

As we pointed out in (ii), the properties of the mass current and angular momentum in the regime of $p_{\rm F}\xi \lesssim 10$ are essentially different from those in the quasiclassical limit. The mass current arising from Weyl-Bogoliubov quasiparticles alone weaken with decreasing $p_{\rm F}\xi$, while high energy quasiparticle states with $|E|>\Delta _{\rm A}$ make a dominant contribution to $L_z/N$. 
In Fig.~\ref{fig:lz}, we plot the $T$-dependence of $L_z/N$ for $p_{\rm F}\xi =10$ and $2.0$, where $\Delta _{\rm A}(T)$ is obtained by calculating the gap equation for a spatially uniform ABM state in Eq.~\eqref{eq:abm}. For comparison, we calculate $L_z^{\rm Cross}=\int ({\bm r}\times {\bm j}^{\rm Cross})_z d{\bm r}$, which is obtained from the gradient expansion as~\cite{crossJLTP75}
\begin{align}
{\bm j}^{\rm Cross}={\bm \rho}_{\rm s}{\bm v}_{{\rm s}}
+\frac{\hbar}{4m}\rho_{\rm s\parallel}{\rm curl}\hat{\bm \ell}
-\frac{\hbar}{2m}\rho_{\rm s\parallel}\hat{\bm \ell}(\hat{\bm \ell}\cdot{\rm curl}\hat{\bm \ell}),
\label{eq:jCross}
\end{align}
where $\rho_{\rm s\parallel}$ and $\rho_{\rm s\perp}$ are the components of the superfluid mass density tensor (${\bm \rho}_{\rm s}$) parallel and perpendicular to $\hat{\bm \ell}$, respectively, and their $T$-dependencies are determined by the generalized Yosida function.~\cite{crossJLTP75} For the case of ${\bm \nabla}\rho={\bm 0}$, Eq.~\eqref{eq:jCross} coincides with Eq.~\eqref{eq:currentMM} at $T=0$, ${\bm j}^{\rm Cross}(T=0)={\bm j}^{\rm MM}$. The averaged superfluid density, $\rho _{\rm ave}\equiv(\rho _{\perp}+\rho _{\parallel})/2$, describes the $T$-dependence of $L^{\rm Cross}(T)$ for $p_{\rm F}\xi\gg 1$, i.e., $L_z^{\rm Cross}(T)/L_z^{\rm Cross}(0)=\rho _{\rm ave}(T)$.
 
As shown in Fig.~\ref{fig:lz}(a), the $T$-dependence of $L_z(T)/L_z(0)$ is in good agreement with that of $L_z^{\rm Cross}(T)/L_z^{\rm Cross}(0)=\rho _{\rm ave}(T)$  in the weak coupling regime, $p_{\rm F}\xi=10$. 
As the topological phase transition ($p_{\rm F}\xi = 0$) is approached, however, the $T$-dependence of $L_z(T)/L_z(0)$ is distinct from that of $\rho _{\rm ave}(T)$. In Figs.~\ref{fig:lz}(c) and \ref{fig:lz}(d), we plot $L_z^{\rm Weyl}(T)/L_z^{\rm Weyl}(0)$, the contribution of the Weyl-Bogoliubov quasiparticles to the angular momentum. In $p_{\rm F}\xi =10$, the $T$-dependence is slightly deviated from that of $\rho _{\rm ave}$ and enhanced at the low $T$ regime. This is understandable with an extra contribution of the torsional chiral magnetic effect, $L^{\rm TCME}_z$, as discussed in Fig.~\ref{fig:lz0}. Such extra contribution vanishes as the topological phase transition ($p_{\rm F}\xi =0$) is approached and the mass current is dominated by the contribution arising from continuum states. 

\begin{figure}[t!]
\includegraphics[width=85mm]{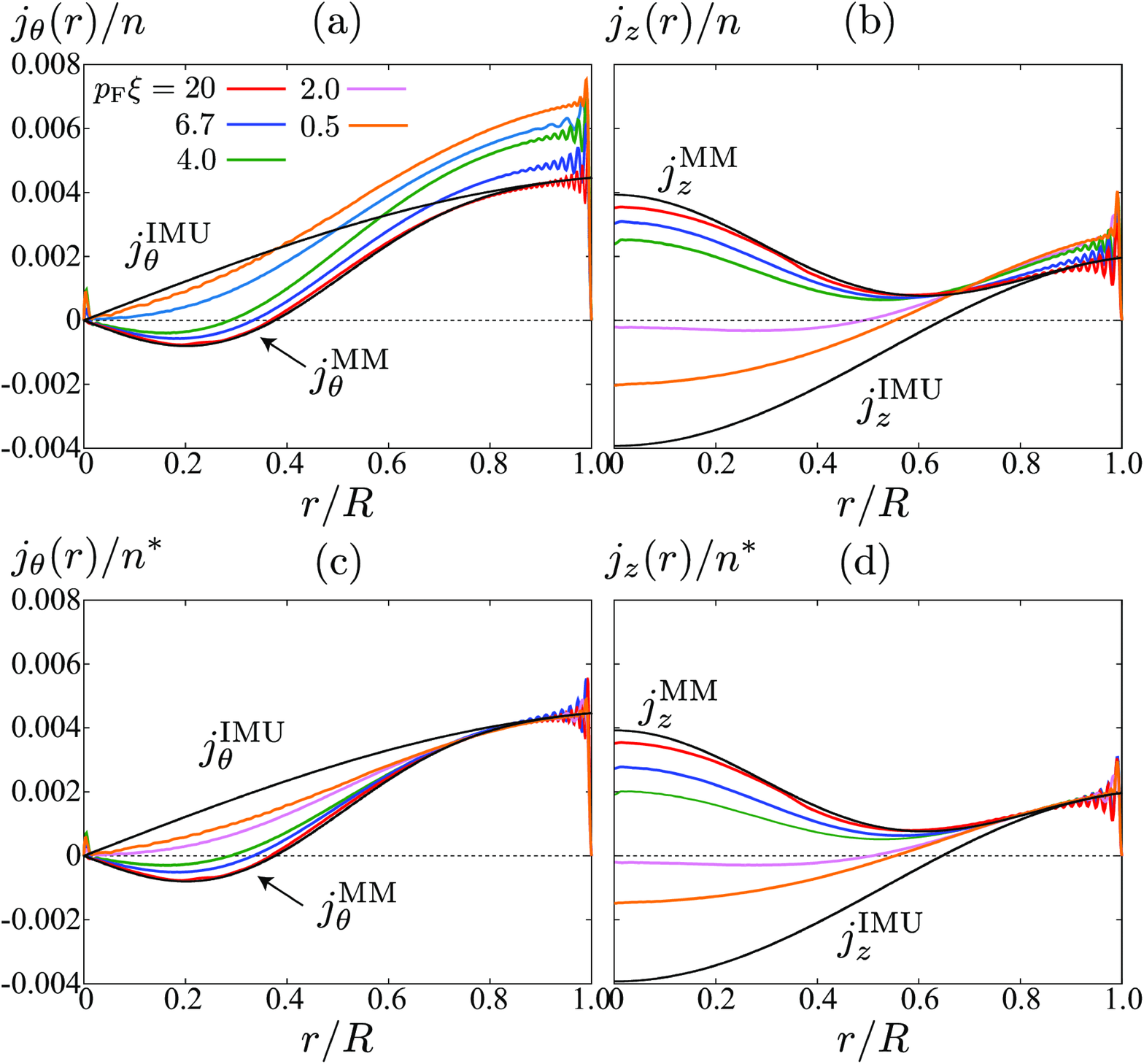}
\caption{Current densities in the twisted skyrmion-vortex state at $T=0$ and $R=200k^{-1}_{\rm F}$: (a) azimuthal current $j_{\theta}(r)/n$ and (b) axial current $j_z(r)/n$. The particle number density $n$ is obtained from the eigenstates of the BdG equation. (c,d) Rescaled current densities, where $n^{\ast}$ is a fitting parameter to rescale $j_{\mu}(r)$ to $j^{\rm MM}_{\mu}(R)=j^{\rm IMU}_{\mu}(R)$.}
\label{fig:twist}
\end{figure}

\subsection{Bloch-type skyrmion}

{In twisted textures with $\hat{\bm \ell}\cdot{\rm curl}\hat{\bm \ell}\neq 0$, the existence of the anomalous term, which is the third term in Eq.~\eqref{eq:currentMM}, has been a long-standing issue. The derivation of Eq.~\eqref{eq:currentMM} is based on the configuration-space form of the BCS variational ground state wavefunction.\cite{merminPRB80} The similar approach was also used by Ishikawa, Miyake, and Usui but came to the different conclusion that the anomalous term, ${\bm j}^{\rm an}\propto\hat{\bm \ell}(\hat{\bm \ell}\cdot{\rm curl}\hat{\bm \ell})$, is absent and the resulting mass current is given by ${\bm j}^{\rm IMU}={\bm j}^{\rm MM}-{\bm j}^{\rm an}$.~\cite{ishikawaPTP80} As ${\bm j}^{\rm an}$ violates the McClure-Takagi relation, the discrepancy between ${\bm j}^{\rm MM}$ and ${\bm j}^{\rm IMU}$ is referred to as the McClure-Takagi paradox.~\cite{merminPRB80,kitaJPSJ96,tsuruta} The physical origin of ${\bm j}^{\rm an}$ was addressed by Combescot and Dombre~\cite{combescotPRB83,combescotPRB86} and Balatsky {\it et al.},~\cite{volovik86,volovik86-2,balatskii87} where the latter unveiled the chiral-anomaly aspect of ${\bm j}^{\rm an}$. Solving the BdG equation in Bloch-type skyrmions, we show below that the mass current is well describable with ${\bm j}^{\rm MM}$ in the weak coupling limit, while it approaches ${\bm j}^{\rm IMU}$ as the topological phase transition is approached.}

Here we consider the mass current induced by Bloch-type twisted-skyrmion textures with $\alpha (r)=\frac{\pi}{2}(1-r/R)$. Using the particle-hole symmetry, one obtains the current along the axial direction, $j_z(r)$, from Eq.~\eqref{eq:current} as 
\beq
j_z(r) = 2 \sum _{E<0}k|u_{nmk}(r)|^2,
\eeq
%
at $T=0$. As shown in Fig.~\ref{fig:weyl}(b), the Bloch-type skyrmion with the broken $P_2$ symmetry induces the chiral fermion branches along axial momentum. As the branch has the negative group velocity with respect to $k$ and the $k>0$ region is occupied at $T=0$, the asymmetry branch makes a positive contribution to $j_z(r)$. In Fig.~\ref{fig:twist}, we plot the azimuthal and axial currents, $j_{\theta}(r)$ and $j_z(r)$. In the weak coupling regime ($p_{\rm F}\xi = 20$), both the current profiles are in good agreement with ${\bm j}^{\rm MM}$ including the anomalous term, rather than ${\bm j}^{\rm IMU}$. {The total angular momentum per particle is estimated as $L_z(0)/\hbar N=\{0.4490, 0.4361,0.4339\}$ for $p_{\rm F}\xi = \{10, 20, 40\}$, respectively. As $p_{\rm F}\xi$ increases, the values approach $L_z/\hbar N=0.4336$ obtained from ${\bm j}^{\rm MM}$, but deviate from the McClure-Takagi prediction that $L_z(0)/\hbar N=0.5$ is independent of the $\ell$-texture.~\cite{MT} The depletion of $L_z(0)$ from $\hbar N/2$ in the weak coupling regime is consistent with the predictions in Refs.~\onlinecite{volovik86,volovik86-2,balatskii87} that in the Bloch-type twisted skyrmion, the anomalous current arising from the chiral anomaly of Weyl-Bogoliubov quasiparticles makes extra contribution to ${\bm j}$ when $\mu > 0$. 
We note that in contrast to ${\bm j}^{\rm an}$, the torsional contribution, ${\bm j}^{\rm TCME}$, does not give rise to significant deviation of $L_z/N$ in $p_{\rm F}\xi \rightarrow \infty$, as ${\bm j}^{\rm TCME} \propto 1/(p_{\rm F}\xi)$.}

{It is shown in Fig.~\ref{fig:twist} that the current density deviates from ${\bm j}^{\rm MM}$ as the topological phase transition ($p_{\rm F}\xi=0$) is approached. To capture the change of the spatial profile in $j_{\theta}$ and $j_z$,  we plot in Figs.~\ref{fig:twist}(c) and \ref{fig:twist}(d) the rescaled mass current densities, where $n^{\ast}$ is a fitting parameter to rescale $j_{\mu}(r)$ to $j^{\rm MM}_{\mu}(R)=j^{\rm IMU}_{\mu}(R)$. These figures show that the rescaled profiles gradually shift from ${\bm j}^{\rm MM}$ to ${\bm j}^{\rm IMU}$ as $p_{\rm F}\xi$ decreases. This implies that while the contributions of Weyl-Bogoliubov quasiparticles through ${\bm j}^{\rm an}$ become significant in the weak coupling regime, ${\bm j}^{\rm an}$ and ${\bm j}^{\rm TCME}$ become negligible around $p_{\rm F}\xi \sim 0$.}



\section{Summary}
\label{sec:summary}

{We have investigated chiral anomaly phenomena induced by skyrmion-like $\ell$-textures in the superfluid $^3$He-A which is a prototype of Weyl superfluids. Using the semiclassical theory, we have shown the torsional chiral magnetic effect that a torsion field induced by skyrmion-like $\ell$-textures results in an equilibrium mass current. In general, the texture of the $\ell$-field generates two different emergent fields directly acting on the chirality of Weyl-Bogoliubov quasiparticles: a chiral gauge field (${\bm A}\propto {\rm curl}\hat{\bm \ell}$) and a torsion field ($T_{\mu\nu}^{\bar{a}}$). In the case of a twisted (Bloch-type skyrmion) $\ell$-texture with $\hat{\ell}\cdot{\rm curl}\hat{\bm \ell}\neq 0$, the chiral gauge field leads to the extra mass current along $\hat{\ell}$, which is the third term of Eq.~\eqref{eq:currentMM}  referred to as the anomalous current.\cite{combescotPRB83,combescotPRB86,volovik81,volovik84,volovik85,volovik86,volovik86-2,balatskii87} Here we find the torsional contribution to the mass current. In contrast to the anomalous current, the torsion-induced mass current flows along ${\rm curl}\hat{\bm \ell}$ and exists even in the case of N\'{e}el-type skyrmion with $\hat{\bm \ell}\cdot{\rm curl}\hat{\bm \ell}= 0$.}

{Using the full quantum mechanical BdG equation, we have demonstrated that in skyrmion vortices a chiral fermion branch with spectral asymmetry appears in the low-lying quasiparticle spectrum. The chiral fermion states are responsible for the equilibrium mass flow. Our numerical results, however, show that the total mass current and the total angular momentum differ from those arising from the chiral fermions alone by a factor of $1/2$. The discrepancy is compensated by the backflow arising from the continuum states, and for {N\'{e}el-type} skyrmion vortices, our numerical results in the quasiclassical limit coincide with the prediction by McClure and Takagi,~\cite{MT} $L_z=\hbar N/2$. Furthermore, it has been demonstrated that the angular momentum associated with the Weyl-Bogoliubov quasiparticles increase as the topological phase transition ($p_{\rm F}\xi =0$) is approached. This anomalous behavior is understandable with the torsional contribution of the current due to the torsional chiral magnetic effect in Eq.~\eqref{eq:jTCME}. We have clarified the torsional-anomaly aspect of the mass current density in Eq.~\eqref{eq:currentMM}; the ${\rm curl}\hat{\bm \ell}$ term in Eq.~\eqref{eq:currentMM} is associated with the TCME of Weyl-Bogoliubov quasiparticles induced by a skyrmion-vortex.}

The appearance of a chiral branch in $^3$He-A was pointed out by Combescot and Dombre,~\cite{combescotPRB83,combescotPRB86} who demonstrated that in the case of a twisted (non-skyrmionic) $\ell$-texture ($\hat{\bm \ell}\parallel {\rm curl}\hat{\bm \ell}$) the BdG equation for low-lying quasiparticles reduces to the Dirac-type equation with a fictitious magnetic field generated by a variation of the $\ell$-field. Balatsky {\it et al.} clarified that the chiral branch is topologically protected by the Atiyah-Singer index theorem and the chiral fermion carries uncompensated current at $T=0$.\cite{balatskii87} Although our result for the mass current qualitatively agrees with that in Ref.~\onlinecite{balatskii87}, it differs from Ref.~\onlinecite{balatskii87} because they consider only the chiral fermion contributions. As mentioned above, we find that in the quasiclassical limit, the continuum states bring about backflow to the quasiparticle flow, i.e., $L_z^{\rm Weyl}\approx - 2L_z^{\rm cont} \approx N\hbar$. As the topological phase transition is approached, the mass current carried by the continuum states changes its sign and makes a dominant contribution. In the vicinity of the topological phase transition ($p_{\rm F}\xi = 0$), indeed, the total angular momentum is governed by the continuum states and the contribution from chiral fermions is negligible. We have also shown that the contribution of ${\bm j}^{\rm an}$ is crucial for the weak coupling regime, while it vanishes as $p_{\rm F}\xi$ decreases. Although $L_z/N$ is composed of the composite contributions of Weyl quasiparticles and continuum states and it is difficult to extract the TCME contribution solely, our results may put a different aspect on the paradox of the mass current and the intrinsic angular momentum.

\begin{acknowledgments}
This work was supported by the Grant-in-Aids for Scientific
Research from MEXT of Japan [Grants No. 23540406, No. 25220711, No. JP17K05517, No. JP16K05448, No. JP15H05852, and No. JP15H05855 (KAKENHI on Innovative Areas "Topological Materials Science"), and No. JP18H04318]. 
\end{acknowledgments}

\bibliographystyle{apsrev4-1_PRX_style} 
\bibliography{WSC}


\end{document}